\documentclass[journal]{IEEEtran} 
\IEEEoverridecommandlockouts
\usepackage{cite}
\usepackage{CJKutf8}
\usepackage{amsmath,amssymb,amsfonts}
\usepackage{algorithmic}
\usepackage{graphicx}
\usepackage{array}
\graphicspath{ {./images/} }
\usepackage{caption}
\usepackage[justification=centering]{caption}
\usepackage[font=small]{caption} 
\usepackage{float}
\usepackage{textcomp}
\usepackage{xcolor}
\usepackage{multirow}
\usepackage{booktabs}
\usepackage{geometry}
\def\BibTeX{{\rm B\kern-.05em{\sc i\kern-.025em b}\kern-.08em
    T\kern-.1667em\lower.7ex\hbox{E}\kern-.125emX}}
\begin{document}

\title{An Enhanced Dynamic Ray Tracing Architecture for Channel Prediction Based on Multipath Bidirectional Geometry and Field Extrapolation\\

}

\author{
     \IEEEauthorblockN{\normalsize Yinghe Miao}, \IEEEauthorblockN{\normalsize Li Yu}, \IEEEauthorblockN{\normalsize Yuxiang Zhang}, \IEEEauthorblockN{\normalsize Hongbo Xing}, \IEEEauthorblockN{\normalsize Jianhua Zhang}
    \\
    \IEEEauthorblockA{\small State Key Lab of Networking and Switching Technology, Beijing University of Posts and Telecommunications, Beijing, 100876, China}
    \IEEEauthorblockA{\small\{yhmiao, li.yu, zhangyx, hbxing, jhzhang\} @bupt.edu.cn}\vspace{-1.0em}
}

\maketitle
\pagestyle{empty}  
\thispagestyle{empty} 

\begin{abstract}
With the development of sixth generation (6G) networks toward digitalization and intelligentization of communications, rapid and precise channel prediction is crucial for the network potential release. Interestingly, a dynamic ray tracing (DRT) approach for channel prediction has recently been proposed, which utilizes the results of traditional RT to extrapolate the multipath geometry evolution. However, both the priori environmental data and the regularity in multipath evolution can be further utilized. In this work, an enhanced-dynamic ray tracing (E-DRT) algorithm architecture based on multipath bidirectional extrapolation has been proposed. In terms of accuracy, all available environment information is utilized to predict the birth and death processes of multipath components (MPCs) through bidirectional geometry extrapolation. In terms of efficiency, bidirectional electric field extrapolation is employed based on the evolution regularity of the MPCs' electric field. The results in a Vehicle-to-Vehicle (V2V) scenario show that E-DRT improves the accuracy of the channel prediction from 68.3\% to 94.8\% while reducing the runtime by 7.2\% compared to DRT.

\end{abstract}


\section{Introduction}

To meet the high quality and full coverage requirements of future communication services, various enabling technologies, such as digital twin \cite{b1} and integrated sensing and communication \cite{b2}, will be adopted in 6G. The network faces the great challenge of working efficiently and flexibly in a more complex and diverse propagation environment. To unlock more potential, accurately acquiring the propagation channel in advance for changing environments becomes crucial, significantly influencing further applications\cite{b3}. Leveraging the advancements in artificial intelligence (AI), AI-based techniques for channel prediction have been extensively explored\cite{b4}, \cite{b5}. However, the challenge of generalization persists in AI-based channel prediction.

Interestingly, in recent years, a dynamic ray tracing (DRT) algorithm has been proposed \cite{b6}, which could be used to predict multipath channel in changing environments with a potentially stronger generalization. Based on the results of traditional RT and the principle of ray-optic propagation, DRT extrapolates the geometry evolution of MPCs, enabling analytical or numerical prediction \cite{b7}. In \cite{b8}, assisted by a complete dynamic environmental database, DRT has been applied to channel prediction. Furthermore, algorithms for extrapolating reflection, diffraction, and diffuse scattering have been developed. The multipath channel at subsequent moments can be predicted by DRT, with a substantial reduction in computational time compared to RT simulation.

However, there is still potential room for improvement in both accuracy and efficiency in prediction. DRT assumes the multipath structure remains static during prediction. However, the birth and death of major MPCs frequently occur in the real world, particularly in a vehicular environment, which leads to unignorable changes in the multipath structure. Considering the birth and death of MPCs can enhance prediction performance. Moreover, the evolution of the electric field on the same MPC at different times can be utilized to extrapolate MPCs' electric field in the prediction. This method may reduce runtime compared to direct electric field computation.


In this work, we propose an enhanced-dynamic ray tracing (E-DRT) algorithm architecture that demonstrates superior accuracy and efficiency compared to DRT for channel prediction. E-DRT leverages all available environment information to predict the processes of MPCs' birth and death furthest through bidirectional geometry extrapolation. Concurrently, bidirectional electric field extrapolation is employed by E-DRT based on the evolution of the MPCs' electric field. It improves computational efficiency while enables the acquisition of all MPCs' parameters through predictive methods. The performance enhancement of E-DRT has been validated through the simulation of a urban dynamic scenario.

The rest of the paper is organized as follows. Sections II and III introduce the DRT and E-DRT prediction algorithms, respectively. The simulation and result analysis are presented in Section IV. Conclusions are drawn in Section V.

\section{DRT algorithm for channel prediction}
DRT is a new paradigm for RT tailored to dynamic scenarios. It utilizes complete environment information, including the dynamic environment and transceiver, to predict the geometric evolution process of MPCs. Based on a traditional RT run at a reference time $t_0$, DRT calculates how each MPCs' interaction points move within the time interval from $t_0$ to $t_0+\mathrm{T_c}$. The $\mathrm{T_c}$ is the extrapolation time determined by the channel coherence time or cluster lifetime. This approach enables the prediction of the geometric structure of MPCs at any time $t_0+t_i$ (where $t_i < \mathrm{T_c}$). It replaces the path-finding process of RT, significantly enhancing computational efficiency.

In DRT, the process commences with a traditional RT run at reference time $t_0$. Then, the position of all MPCs' interaction points are extracted. It is generally assumed that the entire motion of the scatterers and the transceiver is a priori known within ${\mathrm{T_c}}$ \cite{b8}. For example, in \cite{b8}, the velocity $\overrightarrow{v}(t_0)$ and acceleration $\overrightarrow{a}(t_0)$ are assumed to be known at $t_0$, their position $\overline{r}(t)$ at any given time $t_0+t_i$ is
\begin{equation}
\begin{aligned}\overrightarrow{r}(t_0+t_i)=\overrightarrow{r}(t_0)+\overrightarrow{v}(t_0)t_i& +\frac{1}{2}\overrightarrow{a}(t_0)t_i^2+O(t_i^3).\\ \end{aligned}
\end{equation}

Next, such a comprehensive data and the ray-optic propagation principle are utilized. Closed-form formulas for the motion trajectory of the MPCs' interaction points are calculated, including position, velocity, and acceleration \cite{b8}. Then, DRT predicts the geometric evolution of MPCs directly at any $t_i$ within ${\mathrm{T_c}}$ based on trajectory equations. Fig. 1 illustrates an example of the geometry prediction of reflection. Finally, DRT employs the electric field computation step of traditional RT to obtain MPCs' electric field. 

When $t_i > \mathrm{T_c}$, RT should be rerun to predict the MPCs within the next $\mathrm{T_c}$ based on the updated multipath structure. The accuracy of this prediction is assured under the condition that the multipath structure remains constant within $\mathrm{T_c}$, meaning no birth or death of major MPCs.

\begin{figure}[h]
\centering
\includegraphics[width=6.8cm]{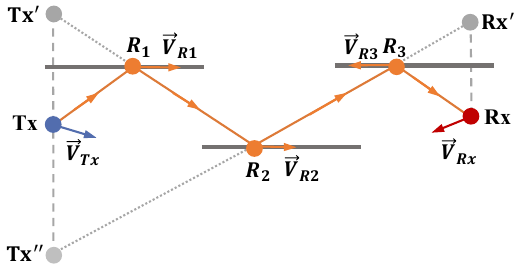}
\captionsetup{justification=raggedright,singlelinecheck=false}
\caption{An example of reflection geometry evolution prediction based on motion information and the ray propagation principle.}
\vspace{-0.5cm}
\end{figure}

\section{The proposed E-DRT prediction algorithm}
\subsection{The Potential Enhancement of DRT}

In terms of accuracy, DRT relies on the assumption of no major MPCs birth or death in prediction. As shown in Fig. 2 (a), the vehicle is in motion, acting as both the transmitter (Tx) and receiver (Rx). The presence of $Ray_{F_1}$ from wall $F_1$ and $Ray_{F_2}$ from wall $F_2$ persists from $t_0$ to $t_0+\mathrm{T_c}$. However, the statistical $\mathrm{T_c}$ cannot guarantee that multipath structure remains constant at all times, especially in vehicle scenarios\cite{b9}. In Fig. 2 (b), solid and dashed lines indicate the presence and absence of MPCs, respectively. Due to the reflection point has already moved away from the wall $F_3$, $Ray_{F_3}$ has died at $t_0+t_i$. Similarly, $Ray_{F_4}$ has been born at $t_0+t_i$ as the reflection point is already on the wall $F_4$. Such phenomena cannot be predicted solely based on the RT of $t_0$. Only when the RT is rerun at $t_0+\mathrm{T_c}$ does DRT discover the death of $Ray_{F_3}$ and the birth of $Ray_{F_4}$. However, these errors have already occurred in the prediction at earlier times like $t_0+t_i$.

\begin{figure}[h]
\centering
\includegraphics[width=8.5cm]{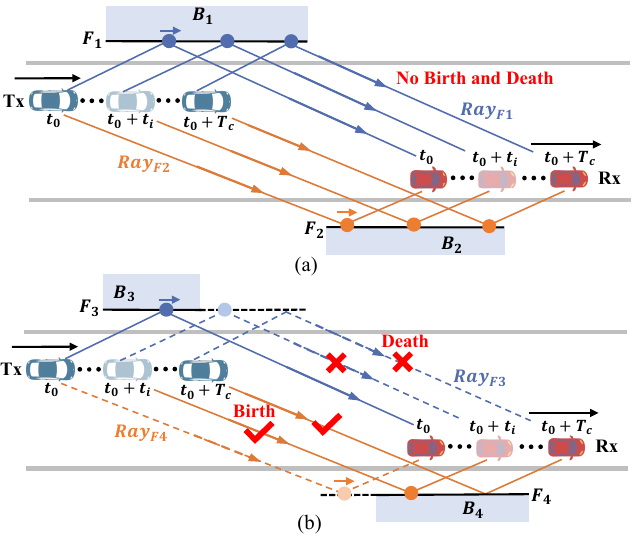}
\captionsetup{justification=raggedright,singlelinecheck=false}
\caption{The two cases of the MPCs geometry evolution in a vehicular environment. (a) represents the assumption of DRT, with no major birth and death of major MPCs within the prediction time $\mathrm{T_c}$. (b) illustrates other case that DRT not consider, where $Ray_{F_3}$ has died and $Ray_{F_4}$ has been born at $t_0+t_i$ ( $t_i < \mathrm{T_c}$).}
\end{figure}

Therefore, the RT results at $t_0+\mathrm{T_c}$, not only enable the next round prediction within $t_0+\mathrm{T_c}$ to $t_0+2\mathrm{T_c}$ but also provide insights into the MPCs' birth and death within $t_0$ to $t_0+\mathrm{T_c}$. If the RT results of $t_0+\mathrm{T_c}$ could be anticipated at $t_0$, the birth and death times of MPCs could be further calculated. This could enable a bidirectional extrapolation of MPCs' geometry evolution. Fortunately, since the motion of all environment objects within $t_0$ to $t_0+\mathrm{T_c}$ is known, the positions of all scatterers and transceiver can be calculated at $t_0$, for instance, using equation (1). Therefore, the RT results of $t_0+\mathrm{T_c}$ can be obtained at the outset of the prediction and improve the accuracy through the bidirectional extrapolation. 

In terms of efficiency, the electric field is still obtained through the traditional electric field calculation step of RT. However, the MPCs that originate from the same scatterers exhibit regularly not only in geometry but also in electric field. Likewise, the relationship of electric field for the same MPC at different times can be derived. This enables direct extrapolation of the MPCs' electric field during prediction, resulting in efficiency gains.

\subsection{The Architecture of E-DRT Algorithm}
E-DRT extends upon the DRT by adding the functionality to predict the birth and death of MPCs. It only advances RT run simply from $t_0+n\mathrm{T_c}$ to $t_0+(n-1)\mathrm{T_c}$, without introducing additional RT runs. The architecture of E-DRT is shown in Fig. 3, with $t_0=0$. First, at time $n\mathrm{T_c}$, E-DRT acquires all motion parameters to calculate scenario information of $(n+1)\mathrm{T_c}$. Then RT is run for time $(n+1)\mathrm{T_c}$. Second, since the RT results of $n\mathrm{T_c}$ have already been computed at $(n-1)\mathrm{T_c}$, the trajectory equations of the MPCs' interaction points are calculated as usual and used in the prediction\cite{b8}. Third, the MPCs that are expected to be born or to die are selected, and their lifetimes are calculated based on the RT results of $n\mathrm{T_c}$ and $(n+1)\mathrm{T_c}$. Then, EDRT predicts the MPCs structure and their birth and deaths from $n\mathrm{T_c}$ to $(n+1)\mathrm{T_c}$. Finally, the electric field extrapolation is conducted. Specifically, at time 0, the RT is run for both time 0 and $\mathrm{T_c}$.

\begin{figure}[h]
\centering
\includegraphics[width=8.5cm]{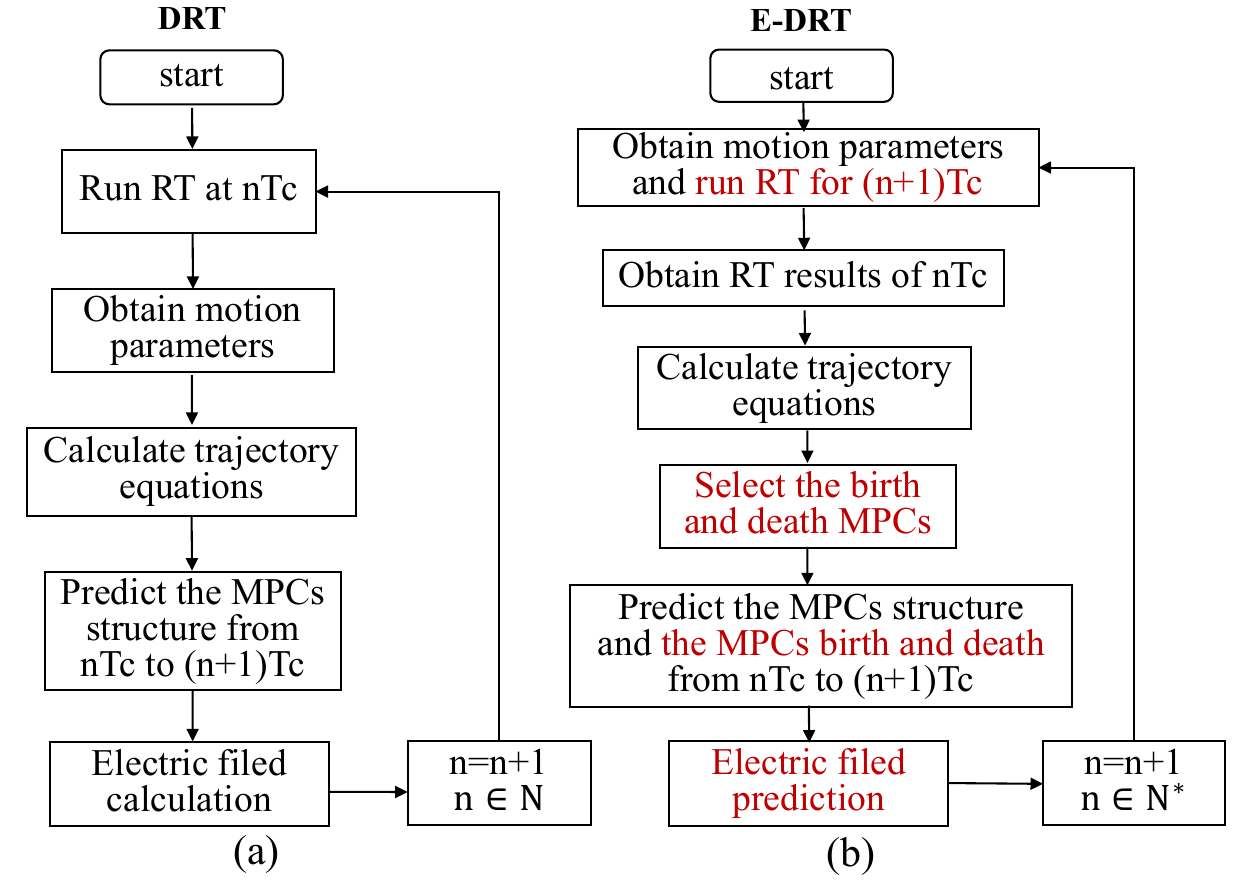}
\captionsetup{justification=raggedright,singlelinecheck=false}
\caption{Comparison of DRT and E-DRT architecture: (a) and (b) represent the process of DRT and E-DRT, respectively, with the start time as the reference time $(t_0=0)$. The $n\mathrm{T_c}$ represents the $n$th prediction round. The process with red characters is the additional process introduced by E-DRT.}
\vspace{-0.5cm}
\end{figure}

\subsection{Selection of Birth and Death MPCs}

Based on the RT results of $n\mathrm{T_c}$ and $(n+1)\mathrm{T_c}$, the similarity of all MPCs is initially compared, including reflection, diffraction, and diffuse scattering. If MPCs originating from the same combination of reflection surface (or diffraction edge) are present at both $n\mathrm{T_c}$ and $(n+1)\mathrm{T_c}$, their mechanism remains unchanged. The disparity in interaction points is solely attributed to the environment changes and transceiver movements. These MPCs are referred to as ``Common Existence MPCs", such as the reflection paths shown in Fig. 2 (a). Conversely, MPCs that appear solely in RT results of $n\mathrm{T_c}$ or $(n+1)\mathrm{T_c}$ are referred to as ``Individual Existence MPCs", such as the reflection path shown in Fig. 2 (b). E-DRT employs different approaches to handle two types of MPCs.
\subsubsection{Common Existence MPCs}

As shown in Fig. 2(a), the $Ray_{F_1}$ from wall $F_1$ and $Ray_{F_2}$ from wall $F_2$ are both common existence MPCs. According to the ray-optic propagation principle and geometric constraint, The interaction points of these MPCs did not depart from their respective surfaces (or edges) at any intermediate time $n\mathrm{T_c}+t_i, t_i < \mathrm{T_c}$. Except for transient blockage or other special circumstances, the lifetime of these MPCs is the entire $\mathrm{T_c}$, with no birth and death.

\subsubsection{Individual Existence MPCs}
If individual existence paths occur, it suggests that the birth and death of MPCs will occur during the prediction from $n\mathrm{T_c}$ to $(n+1)\mathrm{T_c}$. Taking Fig. 2(b) as an example, $Ray_{B_3}$ from building $B_3$ will die, while $Ray_{B_4}$ from building $B_4$ will be born. For MPCs that exist individually, their lifetime is less than $\mathrm{T_c}$.

\subsection{Prediction of MPCs Birth and Death}

To predict potential MPCs birth and death, it is essential to first calculate the lifetimes of selected birth and death MPCs. On one hand, for the $n$th prediction round, MPCs that appear individually at $n\mathrm{T_c}$ will die. Based on the trajectory equation of the MPCs' interaction points and complete environmental data, the solution for when points move to the boundary of their corresponding surfaces can be determined. It represents the final existence time of the interaction points. The accurate death time of this MPC is the minimum value among the final existence times of all interaction points.

\begin{figure}[h]
\vspace{-0.4cm}
\centering
\includegraphics[width=5.8cm]{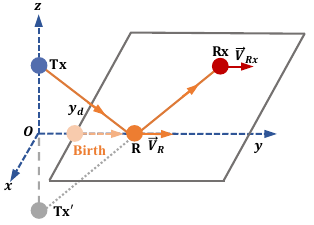}
\captionsetup{justification=raggedright,singlelinecheck=false}
\caption{Example of the birth time calculation of reflection, based on the precomputed RT result of $(n+1)\mathrm{T_c}$.}
\end{figure}
On the other hand, MPCs that appear individually at $(n+1)\mathrm{T_c}$ will be born within the interval from $n\mathrm{T_c}$ to $(n+1)\mathrm{T_c}$. E-DRT utilizes the motion information of the MPCs' interaction points of $(n+1)\mathrm{T_c}$, such as position, velocity, to infer when the MPCs will be born. Without loss of generality, as shown in Fig. 4, a simple 3D reflection case is considered. with a Finite-sized horizontal plane $\Pi: z=0$, the Tx and Rx are located respectively at $(x_{Tx},y_{Tx},z_{Tx})$ and $(x_{Rx},y_{Tx},z_{Tx})$. The equation of lines $\text{Tx}^{\prime}-\text{Rx}$ is
\begin{equation}d:\frac{x-x_{TX^{\prime}}}{x_{RX}-x_{TX^{\prime}}}=\frac{y-y_{TX^{\prime}}}{y_{RX}-y_{TX^{\prime}}}=\frac{z-z_{TX^{\prime}}}{z_{RX}-z_{TX^{\prime}}},\end{equation}
with the Tx image point $\text{Tx}^{\prime}$ given by $(x_{Tx^{\prime}},y_{Tx^{\prime}},z_{Tx^{\prime}})=(x_{Tx},y_{Tx},-z_{Tx})$. Then, the position of reflection point R is the intersection of line $d$ and plane $\Pi$. Deriving the position equation, the velocity equation of R is obtained [7]. 

Assuming that during the time interval from $n\mathrm{T_c}$ to $(n+1)\mathrm{T_c}$, the Tx remains stationary on the z-axis ($z_{Tx}=Z_0$), and the Rx moves uniformly with velocity $\overrightarrow{V_{Rx}}=(0,V_y,0)$ at the same plane with Tx continuously. The position and velocity of R are $\overrightarrow{P_R}=(0,y_{Rx}/2,0)$ and $\overrightarrow{V_R}=(0,V_y/2,0)$ respectively. If the position of Rx at $(n+1)\mathrm{T_c}$ is $(0, Y_{(n+1)\mathrm{T_c}}, Z_0)$, the motion equation of R within the interval from $n\mathrm{T_c}$ to $(n+1)\mathrm{T_c}$ is
\begin{equation}\overrightarrow{P_R}(\Delta t)=\left(0,\frac{Y_{(n+1)\mathrm{T_c}}+\Delta tV_y}2,0\right),\Delta t<0.\end{equation}
If the boundary of plane $\Pi$ is $y=y_d$ in y-axis, then $(n+1)\mathrm{T_c}+\Delta t$ represents the birth time of this MPC, where $\Delta t$ is the solution to the equation $\overrightarrow{P_R}(\Delta t)=(0,y_d,0)$. If the MPC has multiple interaction points, the birth time of that MPC is the the latest birth time of all interaction points.

More computation for the motion trajectory of the MPCs' intersection points in different mechanisms and scenarios has been derived in detail in \cite{b8}. Subsequently, the MPCs' life and death time can be computed. Once E-DRT obtains the lifetime of all MPCs, it can predict the birth and death of MPCs. This process simply involves comparing the prediction time $n\mathrm{T_c}+t_i$ with the birth or death times of the MPCs. Therefore, without introducing additional RT computation points, E-DRT could predict the MPCs structure more accurately through bidirectional geometry extrapolation compared to DRT.
\subsection{Electric Field Extrapolation}
The relationship of electric field for the same MPC at different times is
derived in this subsection. Leveraging this relationship, E-DRT directly extrapolates electric field in the current round prediction, thereby omitting the traditional electric field calculation.

\subsubsection{Reflection Field Extrapolation}

Taking single reflection and the first round of prediction $(n=0)$ as an example, according to the ray propagation theory, the electric field $E_R^{0}$ of Rx at begin time $t_0$ is
\vspace{-0.3cm}
\begin{equation}E_R^{t_0}=E_{t_0}\frac{e^{-jkd_l^{t_0}}}{d_l^{t_0}}\cdot\bar{R}(\overrightarrow{d_l^{t_0}},\overrightarrow{d_p^{t_0}}) \frac{d_l^{t_0}}{d_l^{t_0}+d_p^{t_0}}\cdot e^{-jkd_p^{t_0}},
\vspace{-0.2cm}
\end{equation}
where $k=2\pi/\lambda$ is the propagation constant, $\overrightarrow{d_l^{t_0}}$ and $\overrightarrow{d_p^{t_0}}$ are the incident and  exit vector at the reflection point, $E_{t_0}$ is the emitted electric field of Tx, $E_{t_0}\frac{e^{-jkd_l^{t_0}}}{d_l^{t_0}}$ represents the incident electric field before reflection, $\bar{R}(\overrightarrow{d_l^{t_0}},\overrightarrow{d_p^{t_0}})\frac{d_l^{t_0}}{d_l^{t_0}+d_p^{t_0}}$ denotes the change in direction and magnitude of the electric field caused by reflection, and $e^{-jkd_p^{t_0}}$ represents the phase change due to propagation after reflection. $\bar{R}(\overrightarrow{d_l^{t_0}},\overrightarrow{d_p^{t_0}})$ represents the reflection coefficient for this reflection, which can be calculated using the Fresnel reflection formulas.

Similarly, the received electric field $E_R^{t_i}$ of this reflection path at any prediction time $t_i$ within $\mathrm{T_c}$ is
\vspace{-0.1cm}
\begin{equation}E_R^{t_i}=E_{t_i}\frac{e^{-jkd_l^{t_i}}}{d_l^{t_i}}\cdot\bar{R}(\overrightarrow{d_l^{t_i}},\overrightarrow{d_p^{t_i}})\frac{d_l^{t_i}}{d_l^{t_i}+d_p^{t_i}}\cdot e^{-jkd_p^{t_i}}.\end{equation}
Therefore, the ratio of the electric field magnitude $|E_R^{t_i}|/|E_R^{t_0}|$, which represents the amplitude ratio of the path, can be obtained. Assuming the emitted electric field remains constant, E-DRT directly performs field extrapolation for this reflection field amplitude at the $t_i$ with

\begin{equation}|E_R^{t_i}|=|E_R^{t_0}|\cdot\frac{\bar{R}(\overrightarrow{d_l^{t_i}},\overrightarrow{d_p^{t_i}})}{\bar{R}(\overrightarrow{d_l^{t_0}},\overrightarrow{d_p^{t_0}})}\cdot\frac{d_l^{t_0}+d_p^{t_0}}{d_l^{t_i}+d_p^{t_i}}.\end{equation}
After the geometric prediction is performed by E-DRT, the information of $\overrightarrow{d_l}$ and $\overrightarrow{d_p}$ is complete, allowing for direct electric field extrapolation.
\subsubsection{Diffraction Field Extrapolation}
According to the Uniform Theory of Diffraction (UTD)\cite{b10}, the electric field $E_D^{t_0}$ of the diffraction path at the time $t_0$ is
\begin{equation}
\begin{aligned} 
E_D^{t_0}=E_{t_0}\frac{e^{-jkd_l^{t_0}}}{d_l^{t_0}}\cdot\bar{D}(\overrightarrow{d_l^{t_0}},\overrightarrow{d_p^{t_0}})
\sqrt{\frac{d_l^{t_0}}{(d_l^{t_0}+d_p^{t_0})d_p^{t_0}}}\cdot e^{-jkd_p^{t_0}},
\end{aligned}
\end{equation}
where $\bar{D}(\overrightarrow{d_l^{t_0}},\overrightarrow{d_p^{t_0}})
\sqrt{\frac{d_l^{t_0}}{(d_l^{t_0}+d_p^{t_0})d_p^{t_0}}}$ denotes the change of the electric field caused by diffraction, and $e^{-jkd_p^{t_0}}$ represents the phase change after diffraction. $\bar{D}(\overrightarrow{d_l^{t_0}},\overrightarrow{d_p^{t_0}})$ represents the diffraction coefficient, calculated according to UTD.

Similarly, the electric field $E_D^{t_i}$ of this diffraction path at time $t_i$ can be derived. Finally, The ratio of the path amplitude, $|E_D^{t_i}|/|E_D^{t_0}|$, is obtained. Assuming the emitted electric field remains constant, E-DRT directly perform electric field extrapolation for the diffraction path of the subsequent moment with
\begin{equation}|E_D^{t_i}|=|E_D^{t_0}|\cdot\frac{\bar{D}(\overrightarrow{d_l^{t_i}},\overrightarrow{d_p^{t_i}})}{\bar{D}(\overrightarrow{d_l^{t_0}},\overrightarrow{d_p^{t_0}})}\cdot\sqrt{\frac{(d_l^{t_0}+d_p^{t_0})d_l^{t_i}d_p^{t_0}}{(d_l^{t_i}+d_p^{t_i})d_l^{t_0}d_p^{t_i}}}.\end{equation}
\subsubsection{Transmission and multi-mechanism field extrapolation}
In most cases, transmission coexists with other mechanisms. Taking the example of a reflection path with double penetrations, the relationship between the field amplitude at time 0 and prediction time $t_i$ is 
\begin{equation}
\begin{aligned} 
|E_{R-T}^{t_i}|=|E_{R-T}^{t_0}|\cdot\frac{\bar{R}\left(\overrightarrow{d_{l_R}^{t_i}},\overrightarrow{d_{p_R}^{t_i}}\right)}{\bar{R}\left(\overrightarrow{d_{l_R}^{t_0}},\overrightarrow{d_{p_R}^{t_0}}\right)}\cdot\frac{\bar{T}\left(\overrightarrow{d_{l_T}^{t_i}},\overrightarrow{d_{p_T}^{t_i}}\right)}{\bar{T}\left(\overrightarrow{d_{l_T}^{t_0}},\overrightarrow{d_{p_T}^{t_0}}\right)}\\
\cdot\frac{d_l^{t_0}+d_p^{t_0}}{d_l^{t_i}+d_p^{t_i}}\cdot e^{-\alpha(d_T^{t_i}-d_T^{t_0})},
\end{aligned}
\end{equation}
where $\overrightarrow{d_{l_R}}$ ($\overrightarrow{d_{l_T}}$) and $\overrightarrow{d_{p_T}}$ ($\overrightarrow{d_{p_T}}$) represent the incident and exit vectors at the reflection ( penetration) point, respectively. $\bar{R}$ is the reflection coefficient at the reflection point, and $\bar{T}$ is the transmission coefficient at the penetration point (the sum of the two transmission coefficients). $d_T$ is the penetration distance. $e^{- \alpha d_T}$ represents the additional loss caused by transmission, where $\alpha$ is the attenuation coefficient of the given electromagnetic wave in the penetration material.

Specifically, for MPCs that will die during the prediction, the amplitude of these MPCs is known at time $t_0$. For MPCs that will be born during the prediction, the electric field of these MPCs at time $t_0+\mathrm{T_c}$ is known. Therefore, the field magnitude of all predicted MPCs can be bidirectionally extrapolated using the above equations.

\section{results and analysis}

In this section, the performance of DRT and E-DRT in a real urban scenario is demonstrated and compared based on full RT simulations in multiple dimensions. First, errors in MPCs geometry and electric field prediction are compared, validating the benefits of MPCs' birth and death prediction and the accuracy of electric field extrapolation, respectively. Then, the Power Delay Profile (PDP) and its similarity are compared, representing accuracy in application. Finally, the algorithm complexity is analyzed.
\subsection{Simulation Scenario and Configuration}
To ensure the practical significance of the study, the simulation scenario is selected from the campus of the Beijing University of Posts and Telecommunications in Beijing, China. Real-world Geographic Information System (GIS) data is imported from the OpenStreetMap website \cite{b11}. The GIS data displayed by the QGIS software is illustrated in Fig. 5(a), while the simulation case is demonstrated in Fig. 5(b).
\begin{figure}[h]
\vspace{-0.4cm}
\centering
\captionsetup{justification=raggedright,singlelinecheck=false}
\includegraphics[width=8.5cm]{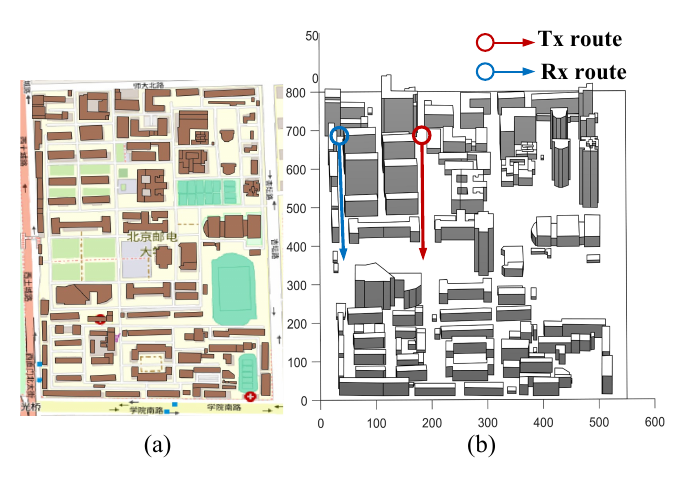}
\caption{The simulation scenario of real urban streets. (a) is the GIS data displayed by the QGSI software and (b) is the simulation case of Vehicle-to-Vehicle (V2V).}
\end{figure}

In this case, which focuses on the Vehicle-to-Vehicle (V2V) scenario, two vehicles serve as the transmitter (Tx) and receiver (Rx) and travel along two parallel routes spanning 285m, with a speed of 36 km/h. The multiple transitions between LOS and NLOS appear in this case, which result in a more frequent change in multipath structure. In RT approach, calculations are performed every 0.1s. For DRT and E-DRT, the $\mathrm{T_c}$ is set to 1s, meaning that after each RT calculation, the next 9 positions (except for the last prediction round) are predicted. The frequency is set at 6 GHz, and both the Tx and Rx adopt omnidirectional antennas. The transmit power of the antennas is 30 dBm. The maximum number of reflection, diffraction, and penetration is set to 2, 1, and 1 respectively.

\subsection{Geometry and Electric Field Prediction Errors of MPCs}

The geometry error $\varepsilon_G$ between prediction and RT simulation is defined in equation (10). $M_p$ represents the number of prediction positions. $N_k^{RT}$ and $N_k^{S}$ respectively represent the number of MPCs in RT and the number of successfully predicted MPCs by DRT/ E-DRT at $k$th position. A MPC is considered successfully predicted when the reflection surfaces or diffraction edges are completely consistent between the prediction and the RT simulation. The electric field error $\varepsilon_E$ is also defined in equation (10), where $E_n^{RT}$ and $E_n^{P}$ denote the field magnitude of the $n$th MPC obtained from RT and prediction, respectively. DRT employs electric field calculation, while E-DRT utilizes electric field extrapolation.
\begin{equation}
\begin{aligned}
\varepsilon_{G}&=\frac{1}{M_{p}}\sum_{k}^{M_{p}}\frac{N_{k}^{RT}-N_{k}^{S}}{N_{k}^{RT}}\\
\varepsilon_{E}&=\frac{1}{M_{p}}\sum_{k}^{M_{p}}\left(\frac{1}{N_{k}^{S}}\sum_{n}^{N_{k}^{S}}\left|\frac{E_{n}^{RT}-E_{n}^{P}}{E_{n}^{RT}}\right|\right)    
\end{aligned}
\label{10}
\end{equation}


\begin{table}[htbp]
\setlength{\tabcolsep}{11pt} 
\renewcommand\arraystretch{1.2}  
\captionsetup{justification=raggedright,singlelinecheck=false}
\caption{The error comparison between DRT and E-DRT. $\varepsilon_{G}$ and $\varepsilon_{E}$ represent the geometry and electric field errors respectively.}
\label{table:I}
\begin{center}
\fontsize{9}{10}\selectfont 
\begin{tabular}{|m{2.8cm}<{\centering}|m{1.2cm}<{\centering}|m{1.2cm}<\centering|}
\hline
 {\text{Prediction Error}} & {\text{DRT}} & {\text{E-DRT}}\\
\hline
 {\(\varepsilon_{G}\times100\%\)} & {$23\%$} & {$4\%$}\\ 
\hline
 {\(\varepsilon_{E}\times100\%\)} & {$0.45\%$}  & {$0.46\%$}\\
\hline
\end{tabular}
\end{center}
\end{table}

Table I illustrates the comparison of prediction errors. The integration of birth and death prediction in E-DRT leads to a significant reduction in geometry error. The mean error of all positions $\varepsilon_{G}$ decreased from 23\% to 4\%. A 4\% error implies that E-DRT cannot guarantee perfect prediction at all times, as certain MPCs may exclusively appear during prediction. Nevertheless, E-DRT maximizes the utilization of available information. Furthermore, the incorporation of electric field extrapolation introduces no additional field errors compared to DRT, and the relative error with concerning to RT simulation remains below 1\%.

\begin{figure*}[t]  
  \centering
  \includegraphics[width=\textwidth]{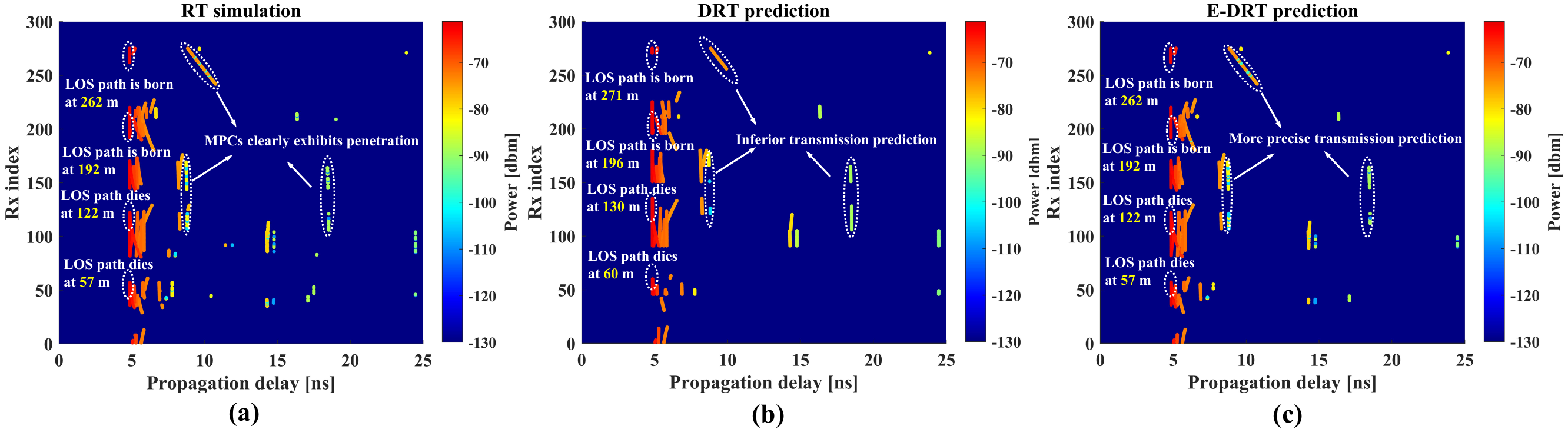}  
  \captionsetup{justification=raggedright,singlelinecheck=false}
  \caption{Comparison of the Power Delay Profiles (PDP) for all positions in a V2V scenario. (a), (b), and (c) denote the major multipath distributions for RT, DRT, and E-DRT, respectively. (c) demonstrates good consistency with RT in terms of MPCs behavior. The birth and death of the LOS path is marked as an example. }
\end{figure*}

\subsection{Power Delay Profile (PDP)} Fig. 6 illustrates the major MPCs distributions for 285 positions, displaying only the first 10 power MPCs. The horizontal axis represents the delay of the MPCs, while the vertical axis represents the Rx movement distance. Fig. 6(a), (b), and (c) denote the PDP of RT, DRT, and E-DRT, respectively. with $\mathrm{T_c}=1\text{s}$, DRT shows relatively accurate prediction for most positions but struggles with positions experiencing abrupt changes in multipath structure. In Fig. 6(b), DRT displays significant errors at the transition between LOS and NLOS, with a maximum error of 9m for the birth and death of the LOS path. Conversely, in Fig. 6(c), E-DRT demonstrates good consistency with RT in terms of MPC behavior, especially in accurately predicting the birth and death of MPCs. Furthermore, to assess the accuracy of the predicted channel, a similarity index is introduced here \cite{b12}. It is expressed as
\begin{equation}SI=1-\frac12\int\left|\frac{{P}^{\mathrm{tar}}(\tau)}{\int{P}^{\mathrm{tar}}(\tau)d\tau}-\frac{{P}^{\mathrm{pre}}(\tau)}{\int{P}^{\mathrm{pre}}(\tau)d\tau}\right|d\tau. \end{equation}
${P}^\mathrm{tar}$ and ${P}^\mathrm{pre}$ denote the power of all MPCs across positions of the target channel and prediction channel, respectively. The range of $SI$ is [0, 1]×100\%, where 100\% denotes full similarity and 0 is maximal dissimilarity. With reference to the RT results, the channel similarity of the DRT is calculated as 68.3\%, and that of E-DRT is 94.8\%.

\subsection{Computational Gain Versus DRT and RT}
To evaluate the computational gain of E-DRT versus DRT and RT, we recorded the calculation time on the 12th Gen Intel (R) Core (MT) i7-12700H @4.7 GHz platform.

\begin{table}[htbp]
\setlength{\tabcolsep}{15pt} 
\renewcommand\arraystretch{1.2}  
\caption{EXECUTION TIME FOR RT, DRT AND E-DRT.}
\label{table:II}
\begin{center}
\fontsize{9}{10}\selectfont 
\begin{tabular}{|m{1.85cm}<{\centering}|m{0.6cm}<{\centering}|m{0.6cm}<\centering|m{1.05cm}<{\centering}|}
\hline
 {\text{Time(s)}} & {\text{RT}} & {\text{DRT}} & {\text{E-DRT}}\\
\hline
  {\text{Geometry}} & {3608} & {339} & {341}\\
\hline
  {\text{Electric field}} & {37} & {36} & {7}\\
\hline
  {\text{Total}} & {3645} & {375} & {348}\\
\hline
\end{tabular}
\end{center}
\end{table}

Observing Table \ref{table:II} reveals that both DRT and E-DRT achieve approximately a 10x computation gain compared to RT, which is associated with the prediction of 9 positions per round. Only a 0.6\% increase in geometry computation time is observed in E-DRT compared to DRT. In addition, the implementation of electric field extrapolation in E-DRT leads to a notable 80\% reduction in computation time compared to DRT. While the computation of electric field itself constitutes a minor portion of the overall prediction process, this enhancement contributes to an overall reduction of 7.2\% in total runtime compared to DRT.

\section{conclusions}
In the present work, an E-DRT algorithm for channel prediction based on multiple bidirectional geometry and field extrapolation has been proposed. Following the implementation of MPCs birth and death prediction, the average geometry error decreased from 23\% to 4\% compared to DRT, with a complexity increase of only 0.6\%. Furthermore, the relative error in electric field extrapolation remains below 1\% while the complexity decreases by 80\%. In summary, E-DRT improves the accuracy of channel prediction in a V2V scenario from 68.3\% to 94.8\% while reducing the runtime by 7.2\%. This work facilitates rapid and precise channel prediction in dynamic scenarios, thereby empowering the development of digital twins for future smart cities, factories, and various other applications.

\section{acknowledgment}
This work is supported by the National Key R\&D Program of China (No. 2023YFB2904803), the National Natural Science Foundation of China (No. 62341128), the National Science Fund for Distinguished Young Scholars (No. 61925102), the National Natural Science Foundation of China (No. 62201087 \& No. 62101069), and BUPT-CMCC Joint Innovation Center.






\vspace{12pt}

\end{document}